\documentclass[11pt,prd,aps,tightenlines,superscriptaddress,nofootinbib]{revtex4}
\usepackage{epsfig}
\usepackage{amsfonts}
\usepackage{amsmath}
\usepackage{graphicx}
\usepackage{color}

\newcommand{\boldgreek}[1]{{\mbox{\boldmath{$#1$}}}}
\newcommand{\bg}{\boldgreek}
\newcommand{\beq}{\begin{eqnarray}}
\newcommand{\eeq}{\end{eqnarray}}

\begin{document}

\title{Energy Dependence of Direct Detection Cross Section for Asymmetric Mirror Dark Matter }
\author{Haipeng An}
\author{Shao-Long Chen}
\author{Rabindra N. Mohapatra}
\affiliation{Maryland Center for Fundamental Physics and Department of Physics, University of Maryland, College
Park, Maryland 20742, USA}
\author{Shmuel Nussinov}
\affiliation{Tel Aviv University, Israel and Chapman Schmid College of Science, Orange, California 92866, USA}
\author{Yue Zhang}
\affiliation{Abdus Salam International Centre for Theoretical
Physics,  Strada Costiera 11, I-34014 Trieste, Italy}

\begin{abstract}
In a recent paper, four of the present authors proposed
a class of dark matter models where generalized parity symmetry
leads to equality of dark matter abundance with baryon
 asymmetry of the Universe and predicts dark
matter mass to be around 5 GeV. In this note we explore how this
model can be tested in direct search experiments. In particular,
we point out that if the dark matter happens to be the mirror
neutron, the direct detection cross section has the unique feature
that it increases at low recoil energy unlike the case of
conventional WIMPs. It is also interesting to note that the
predicted spin-dependent scattering could make significant
contribution to the total direct detection rate, especially for
light nucleus. With this scenario, one could explain recent
DAMA and CoGeNT results.
\end{abstract}

\preprint{\vbox{\hbox{UMD-40762-477}}}
\preprint{\vbox{\hbox{UMD-PP-10-xxx}}}

\maketitle

\section{Introduction}

It is now widely accepted that almost a quarter of the mass-energy
in the Universe is dark matter and one of the major challenges of
particle physics and cosmology is to discover the nature of the
dark matter. Since the standard model of particle physics does not
contain any stable particle that can play the role of dark matter,
this provides evidence for physics beyond standard model (BSM) and
many BSM scenarios have been proposed that include stable or very
long-lived Weakly Interacting Massive Particles (WIMPs) which can
play this role~\cite{Bergstrom:2009ib}. Dark matter being
pervasive in our galaxy with an energy density of $\rho_{DM}\simeq
0.3$\,GeV/cm$^3$, it could be observable by detection of nuclear
recoils produced when it scatters off nuclei in a very
low-background detector~\cite{Goodman:1984dc,Freese:1987wu}. The
recoil energy distribution which is in the keV range could provide
clues to the nature of the WIMP.

Among the direct detection experiments, {\it e.g.},
CDMS~\cite{Ahmed:2008eu} and XENON10~\cite{Angle:2007uj} have not
found any signal from WIMPs and set the most stringent constraints
on the WIMP-nucleon elastic scattering cross section. On the other
hand, DAMA collaboration has reported an annual modulation signal
in the scintillation light from their DAMA/NaI and DAMA/LIBRA
experiments, which is interpreted as evidence of dark
matter~\cite{Bernabei:2003za,Bernabei:2008yi}. The CDMS II
collaboration has observed two possible dark matter signal events
for an expected background of $0.8\pm 0.2$
events~\cite{Ahmed:2009zw}. More recently the CoGeNT collaboration
has published their results from the ultra low noise germanium
detector with a very low energy threshold of $0.4$ keVee in the
Soudan Underground Laboratory~\cite{Aalseth:2010vx}. Although the
observed excess is consistent with an exponential background, it
also could be explained by a WIMP in the mass range $5\sim 10$
GeV, with a rather large WIMP-necleon spin-independent (SI)
elastic scattering cross section $\sim 10^{-40}$
cm$^2$~\cite{Kopp:2009qt,Fitzpatrick:2010em}.

It is well known that the null experiments have already ruled out the case of canonical WIMP masses
$\sim 100$ GeV to be capable of producing the DAMA results. Yet for a low mass
(${\mathcal O}(10)$ GeV) WIMP, the compatibility is possible.

The light dark matter fits of DAMA and CoGeNT motivate many light
DM models; among them, a class of very attractive ones are the
asymmetric dark matter (ADM) models. The ADM models are different
from the usual WIMP models in that whereas the latter have a relic
thermal abundance determined by the thermal `freeze-out', ADM
abundance is related to the baryon asymmetry in the universe
~\cite{Nussinov:1985xr,Kaplan:1991ah,Barr:1991qn,Barr:1990ca,Gudnason:2006ug,Dodelson:1991iv,Fujii:2002aj,Kitano:2004sv,
Kitano:2008tk,Farrar:2005zd,Berezhiani:2008gi,Kaplan:2009ag,Kribs:2009fy,Cai:2009ia,An:2009vq}.

Recently we proposed an ADM model~\cite{An:2009vq} in which the
standard model is accompanied by a dark standard (or mirror) model
which is a complete duplication of the matter and forces in the
visible SM. A mirror symmetry guarantees that prior to symmetry
breaking there are no free coupling parameters in the dark sector.
This is therefore distinct from models where an arbitrary dark
sector is appended to the standard model. Symmetry breaking is
assumed to be different in the mirror sector compared to the
familiar SM sector so that the model is consistent with cosmology.

There are several ways that the two sectors are connected: the first, of course,
is via gravity as every matter would couple to gravity. To
understand small neutrino masses in our sector, we invoke the
seesaw mechanism and add three right-handed neutrinos. A novel
aspect of our model~\cite{An:2009vq} is that instead of adding RH
neutrinos separately to two sectors, we add a common set of three
RH neutrinos that provides a second link between the two
sectors~\cite{bere}. Finally, we add a kinetic mixing between the
$U(1)$ bosons of the two sectors. Other details of the model
are reviewed in Sec.~\ref{sec3}.

The right-handed neutrinos not only help in understanding of the
small neutrino masses by a variation of the usual seesaw
mechanism~\cite{MV}, they also play a crucial role in our
understanding of dark matter abundance: in the early universe, the
RH neutrinos decay out of equilibrium and generate equal leptonic
asymmetry in both sectors. These asymmetries are then transferred
into baryonic and mirror-baryonic asymmetries through the sphaleron
processes in both sectors. Thus the full weak $SU(2)_L$ group in
both sectors are essential to our scenario. The lightest mirror
baryon is considered as the dark matter particle. Thus
baryogenesis via leptogenesis explains both the origin of matter
as well as dark matter, making their number densities equal to each other
due to mirror symmetry. This allows us to predict the dark matter
mass to be $m_N\Omega_{DM}/\Omega_B  \sim 5$ GeV. The
$U(1)-U^\prime(1)$ kinetic mixing  along with a massive mirror photon
helps us to maintain consistency of the model with Big Bang
Nucleosynthesis (with a mirror photon mass in the $10-100$ MeV
range). The mirror photon, therefore,
 provides a portal linking the two sectors and makes the direct
 detection of the dark matter possible. Furthermore, the dark matter in our model has
 self interaction and as pointed out in~\cite{An:2009vq}, the self interaction cross section
 is safely below the bullet cluster constraint.

In this work, we investigate the direct detection of the dark
baryons that arise in the class of asymmetric mirror models
proposed in \cite{An:2009vq}. We write down the general operators
for neutral dark baryon interaction with the visible sector
through a light massive mirror photon portal. We find that the
interactions are energy/momentum dependent and the differential
cross section has non-uniform angular distribution. These new
features are absent in the conventional WIMP case for both
spin-independent (SI) and spin-dependent (SD) interactions. This
provides a way to distinguish between this type of DM from many
familiar DM candidates. We also consider the scenarios when the
charged dark baryon $p^\prime$ or $\Delta^\prime$ is the dark
matter, in which case there is no such momentum dependence.

The paper is organized as follows: in Sec.~\ref{sec2} we give a general operator analysis
of dark matter and nuclear interaction that applies to the asymmetric dark matter and similar models.
In Sec.~\ref{sec3}, we discuss the implications of the general operator analysis
and the energy dependent direct detection cross-section that results for this
general case. In Sec.~\ref{sec4} we present our conclusions.

\section{Direct detection: operator analysis and cross sections}\label{sec2}
Dark matter direct detection experiments measure the recoil energy
deposited when a WIMP collides with a nucleus in the detector. For
a WIMP of mass $m_\chi$ scattering with a nucleus of mass $m_A$,
the recoil energy $E_r$ is given by $E_r=\mu^2 v^2/m_A
(1-\cos\theta)$, where $\mu=m_\chi m_A/(m_\chi+m_A)$ is the reduced
mass and $\theta$ is the scattering angle in the center of mass
frame.

The differential detection rate can be written as
\begin{eqnarray}
\frac{dR}{dE_{r}} = N_T \frac{\rho_0}{m_\chi} \int_{v_{\rm
min}}^{v_{\rm max}} \frac{ d \sigma}{dE_r} v f(v, v_e) d^3{\bf v}\,,
\end{eqnarray}
where $\rho\approx0.3\,{\rm GeV}\,{\rm cm}^{-3}$ is the local DM
density in the solar system, $f(v, v_e)$ is the distribution of DM
velocity and $v$ is the velocity with respect to the Earth, $N_T$ is the number of
scattering nucleus per unit detector mass. For elastic scattering
with given recoil energy $E_r$, the limits of the above integral
are given by $v_{\rm min} = \sqrt{ m_A E_r/(2\mu^2)}$ and $v_{\rm
max}\approx 650$\,km/s, the escape velocity from our galaxy.

The differential cross section induced by DM-nucleus scattering is
given by the spin independent and spin dependent contributions,
which are conventionally written as
\begin{eqnarray}\label{crosssection}
\frac{d\sigma}{dE_r} = \frac{m_A}{2\mu^2 v^2} \left[\sigma_{\rm SI}^0
F^2(|{\bf q}|)+\sigma_{\rm SD}^0 S(|{\bf q}|)/S(0) \rule{0mm}{6mm}\right]\,,
\end{eqnarray}
where $\sigma^0_{\rm SI, SD}$ are the integrated SI and SD DM-nucleus
cross sections. $F(|{\bf q}|)$ is the SI form factor and takes the
common Helm form factor~\cite{Jungman:1995df}
\begin{eqnarray}
F^2(|{\bf q}|) = \left[\frac{3j_1(|{\bf q}| R_1)}{|{\bf q}|
R_1}\right]^2 \exp(-(|{\bf q}| s)^2) \ ,
\end{eqnarray}
where $j_1$ is the first spherical Bessel function, $|{\bf
q}|=\sqrt{2m_AE_r}$, $R_1=(R^2-5s^2)^{1/2}$, $R=1.2\,{\rm fm}\,A^{1/3}$
and $s\approx 1\,$fm. The SD form factor $S(|{\bf q}|)$ is
specific to the target nucleus.

The velocity distribution of DM in the galactic halo is often
assumed to be given by a standard Maxwellian distribution
\beq
f({\bf u})=f({\bf v}+{\bf v}_{e})=\frac{1}{(\pi
v_0^2)^{3/2}}e^{-{\bf u}^2/v_0^2}\,,
\eeq
where $v_0 \sim 270$ km/sec, ${\bf v}$ is the velocity of DM with respect to the detector
and ${\bf v}_e$ is the Earth's speed velocity relative to the halo
and it is time dependent: $v_e= v_\odot +14.4 \cos[2\pi (t-t_0)/T]$
km/sec with $t_0=152$ days and T=1 year. Due to the rotation of
the Earth around the Sun, the direct detection signal for DM has a
well-known annual modulation effect \beq S(E, t)=S_0(E)+A(E) \cos
\left[\frac{2\pi (t-t_0)}{T}\right]\,. \eeq

It is worth pointing out in regard to Eq.~(\ref{crosssection})
that in the discussions so far, the differential cross section is
assumed to be momentum independent except for the nuclear
structure form factor with low momentum transfer. In other words,
the dark matter-nucleon interactions are assumed to be such that
they do not generate momentum dependence in the differential cross
section. However, in general there can be interactions which can
lead to ${\bf q}$-dependence and if the dependence comes with a
large coefficient, it could be detected in laboratory searches.
Examples of DM particles which could lead to such situations are
milli-charged particles or DM particle interacting with dipole
moment. In the model recently proposed by us in~\cite{An:2009vq},
the mirror neutron is considered as the dark matter candidate. It
interacts through a light mirror photon portal. We will show that
the cross sections in this case are ${\bf q}$-dependent and large
enough for direct detection. For other examples of dark matter
models with momentum dependent scattering cross section, see Refs.
~\cite{Sigurdson:2004zp,Foot:2008nw,Alves:2009nf,Masso:2009mu,
Chang:2009yt,Chang:2010yk,Feldstein:2009np,Feldstein:2009tr}.

\subsection{General Operators Analysis}\label{sec21}
In this section, we give a general operator analysis of dark
matter-nuclear interaction where
there is a kinetic mixing $\frac{\varepsilon_\gamma}{2} F^{\mu\nu}F^\prime_{\mu\nu}$
between the mirror sector gauge field and the visible sector, and the
hidden $U(1)^\prime$ is broken so that the mirror photon has a mass
in the range $10-100$ MeV. The light mirror photon becomes the
portal linking the dark matter and SM particles.

The interaction of nucleons with the mirror photon can then be
written as
\begin{eqnarray}
\mathcal{L} = \varepsilon_\gamma e \bar p \gamma^\mu p A_\mu' +
\varepsilon_\gamma \frac{\mu_N}{2} \bar N \sigma^{\mu\nu} N
F_{\mu\nu}' \ ,
\end{eqnarray}
where $N=p, n$ stands for proton and neutron, respectively, and
$\mu_N$ is the anomalous magnetic dipole of the nucleons.

Consider a particle from the mirror sector as the dark matter
candidate, and it carries vanishing mirror
electric charge. Therefore, it interacts with the mirror photon
through its anomalous magnetic dipole moment or other higher
dimensional operators. In analogy to the effective field
theories of nucleons in QCD, we write down all possible operators
up to dimension six.
\begin{eqnarray}\label{Lprime}
\mathcal{L}' &=& c_1 \frac{e}{2m_\chi} \bar \chi \sigma^{\mu\nu}
\chi F_{\mu\nu}' + c_2 \frac{e}{2m_\chi^2} \bar \chi \gamma^\mu
\chi \partial^\nu F_{\mu\nu}' +
 c_3 \frac{e}{m_\chi^2} \bar \chi \gamma^\mu \partial^\nu \chi F_{\mu\nu}'  + {\rm h.c.} \ ,
\end{eqnarray}
where $\mu_\chi= c_1 e/m_\chi$ is defined as the anomalous mirror
magnetic dipole moment of the mirror neutron. It is easy to check
that other operators such as $(e/m_\chi^2)
\varepsilon^{\mu\nu\rho\sigma} \bar \chi \gamma_\mu \gamma_5 \partial_\nu
\chi F_{\rho\sigma}'$ can be decomposed into linear combinations
of the above three.

\begin{figure}[htb]
\begin{center}
\includegraphics[scale=1.4]{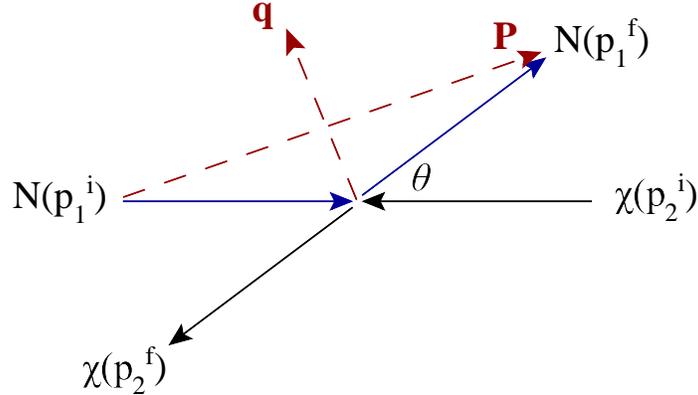}
\caption{The kinematics of scattering: $|{\bf q}|^2 = 2 \mu^2 v^2 (1-\cos\theta)$,
$|{\bf P}|^2 = 2 \mu^2 v^2 (1+\cos\theta)$ and ${\bf q}\cdot {\bf P} = 0$ in the CM frame.}
\label{fk}
\end{center}
\end{figure}

The matrix element of the low-energy scattering between the nucleon and dark matter
 can be obtained by integrating out the mirror photon.
\begin{eqnarray}
\mathcal{M}_{\rm eff} &=& \varepsilon_\gamma \frac{c_1}{m_{\chi}} \frac{e^2}{m_{\gamma '}^2}
(\bar p \gamma^\mu p) q^\nu (\bar \chi \sigma_{\mu\nu} \chi) + i \varepsilon_\gamma \frac{c_1 \mu_N}{m_\chi}
\frac{e}{m_{\gamma '}^2} (\bar N \sigma^{\mu\nu} N)
q_\mu q^\alpha (\bar \chi \sigma_{\alpha\nu} \chi) \nonumber \\
&+& i \varepsilon_\gamma \frac{c_2}{2m_{\chi}^2}
\frac{e^2}{m_{\gamma '}^2} (\bar p \gamma^\mu p) q^2 (\bar \chi
\gamma_\mu \chi) + i \varepsilon_\gamma \frac{c_3}{m_\chi^2}
 \frac{e^2}{m_{\gamma '}^2} (\bar p \gamma^\mu p)
q^\nu \left[\bar\chi ( \gamma_\mu P_\nu - \gamma_\nu P_\mu ) \chi \right] ,
\end{eqnarray}
where $q$ is the momentum transfer and $P$ is the sum of momenta of the initial and final
nucleons. The kinematics of scattering is shown in Fig.~\ref{fk}. In the center of mass
 (CM) frame, one has $P^0 \sim 2\mu$,
$q^0 \sim \mathcal{O}(\mu v^2)$, the three-momenta $P^i, q^i \sim
\mu v$ satisfying ${\bf q}\cdot {\bf P}=0$, $\mu$ is the reduced
mass and $v$ is the velocity of the incoming dark matter particle in
the laboratory frame. Based on the power counting, we perform a
nonrelativistic reduction of the above operators\footnote[1]{We
choose the following representation
\begin{equation*}
  \gamma^0  = \left[\begin{array}{cc} 1 & 0 \\ 0 & -1 \end{array}\right], \, \, \, \, \,
\gamma^i = \left[\begin{array}{cc} 0 & \sigma^i \\ -\sigma^i & 0
\end{array}\right], \, \, \, \, \, N(p) =
\sqrt{\frac{p^0+m_N}{2p^0}} \left[\begin{array}{c}N_h \\
\frac{\mathbf{p} \cdot{\bg \sigma}}{p^0+m_N}N_h \end{array}\right]
\end{equation*}
where $\sigma^i$ is the Pauli matrix and $N_h$ is the nonrelativistic two-component nucleon field.
The dark matter has a similar form, with the nonrelativisitic field denoted as $\chi_h$.}.
The nonrelativistic reduction of the scattering amplitude yields
\begin{eqnarray}\label{Anr}
\mathcal{M}_{\rm nr}
&=& \varepsilon_\gamma \frac{(c_1+c_2) e^2}{2m_\chi^2 m_{\gamma '}^2} |{\bf q}|^2 (p_h^{\dag} p_h ) (
\chi_h^{\dag} \chi_h ) +
 \varepsilon_\gamma \frac{ c_1 e^2}{2\mu m_\chi m_{\gamma '}^2} ( {\bf q} \times {\bf P} )^i ( p_h^{\dag} p_h )
( \chi_h^{\dag} \sigma^i \chi_h ) \nonumber \\
&+& \varepsilon_\gamma \frac{(\frac{e}{2m_p}+\mu_p) c_1 e}{m_\chi m_{\gamma '}^2} (|{\bf q}|^2 \delta_{ij} - q^i q^j)
( p_h^{\dag} \sigma^i p_h ) ( \chi^\dag \sigma^j \chi_h) \nonumber \\
&+& \varepsilon_\gamma \frac{\mu_n c_1 e}{m_\chi m_{\gamma '}^2} (|{\bf q}|^2 \delta_{ij} - q^i q^j)
( n_h^{\dag} \sigma^i n_h ) ( \chi^\dag \sigma^j \chi_h) \ .
\end{eqnarray}
where $p_h (n_h),  \chi_h$ are the non-relativistic two-component nucleon and dark matter fields respectively.

Several comments are in order.
\begin{itemize}
\item The higher dimensional operators in Eq.~(\ref{Lprime}) are parity even,
which differ from those considered in~\cite{Chang:2009yt}.
We write down the operators up to dimension six in the mirror sector. The dimension six operator
($c_2$ term in Eq.~(\ref{Lprime}))
is relevant for the completeness of studying the momentum dependent direct detection,
 since it contributes in the same order as the magnetic dipole, as shown in
Eq.~(\ref{Anr}).

\item It is interesting to note that each term in the amplitude
$\mathcal{M}_{\rm nr}$ is ${\bf q}$-dependent, and proportional to
(suppressed by) the momentum power $|{\bf q}|^2$ or $( {\bf q}
\times {\bf P} )\cdot{\bg \sigma}$. The dark matter-nucleon
scattering happens at higher partial waves ($\ell>0$) instead of
s-wave. This is due to the fact that the dark matter is
electrically neutral. Thus the leading order interaction includes
those between the proton electric charge and the dark matter
magnetic dipole or two dipoles. In the denominator, we also have a
small mass scale $m_{\gamma'}^2\approx (10-100\,{\rm MeV})^2$ , which
will compensate for the suppression in the numerator.

\item The nucleon part of each term has the form of either $p_h^\dag p_h$ or $N_h^\dag \sigma^i N_h$,
corresponding to spin-independent and spin-dependent cross sections, respectively. No new form factor is needed
 for calculating the nuclear-level cross sections.

\item  The $c_3$ term does not contribute to the
amplitude. The reason is that after nonrelativistic reduction,
the corresponding operator results in either ${\bf q}\cdot{\bf P}$
or ${\bf q}\times{\bf q}$ and both vanish.
\end{itemize}

For the scenario when the mirror charged baryon is chosen as the DM candidate, as
we will show in Sec.~\ref{sec_charged},
an additional term $c_0 e \bar \chi \gamma^\mu \chi A_\mu' $ will be added into Eq.~(\ref{Lprime}). Therefore
in the nonrelativistic limit the operator $\bar \chi \chi \bar pp$ with zero-th power of $|{\bf q}^2|$ will
dominate the interaction and
it is the conventional SI type interaction.

\subsection{General ${\bf q}$-Dependent Cross Sections}

The SI differential cross section induced by dark matter-nucleon effective interactions is
\begin{eqnarray}\label{SI0}
\frac{d \sigma_{\rm SI}}{dE_r} = \varepsilon_\gamma^2 \frac{Z^2
e^4 m_A \mu^2 v^2}{2\pi m_\chi^2m_{\gamma'}^4} \left[ (c_1+c_2)^2\frac{\mu^2}{m_\chi^2}
(1-\cos\theta)^2+ c_1^2 \sin^2 \theta\right]
F^2(|{\bf q}|)\,,
\end{eqnarray}
where $\theta$ is related to
$|{\bf q}|^2$ and $v$ via $|{\bf q}|^2 = 2 \mu^2 v^2 (1-\cos\theta)$.

The SD part of differential cross sections is
\begin{eqnarray}\label{SD0}
\frac{d \sigma_{\rm SD}}{dE_r} =
\frac{m_A |{\bf q}|^4}{3\pi  v^2 m_{\gamma'}^4}
\left[ a_0^2 S_{00}(0) + a_0 a_1 S_{01}(0) + a_1^2 S_{11}(0) \rule{0cm}{6mm}\right]
\frac{J_A+1}{J_A} \frac{S(|{\bf q}|)}{S(0)} \ .
\end{eqnarray}
For the SD form factor, we will choose that given in~\cite{Belanger:2008sj},
\begin{eqnarray}
\frac{S(|{\bf q}|)}{S(0)} = \exp (- |{\bf q}|^2 R_A^2/4) \ ,
\end{eqnarray}
where $R_A = 1.7 A^{1/3} - 0.28 - 0.78( A^{1/3} - 3.8 + \sqrt{ (A^{1/3}-3.8)^2 + 0.2 })$ fm,
and $S_{00}(0)=(S^A_p+S^A_n)^2$, $S_{11}(0)=(S^A_p-S^A_n)^2$
and $S_{01}(0)=2 (S^A_p+S^A_n)(S^A_p-S^A_n)$, where $S_N^A\approx0.5$ ($N=p,n$) or 0
for a nucleus containing odd or even number of nucleons $N$.
The isoscalar and isovector part coefficients are $a_0 = \xi_p+\xi_n$ and $a_1 = \xi_p-\xi_n$,
where $\xi_N$ are defined here as
\begin{eqnarray}
\xi_p = \varepsilon_\gamma \left(\frac{e}{2m_p}+\mu_p\right) \mu_\chi , \, \,
\xi_n = \varepsilon_\gamma \mu_n \mu_\chi \ ,
\end{eqnarray}
with $\mu_\chi \equiv c_1 e/m_\chi$. When $\mu_\chi\neq0$, there are always both SI and SD contributions.

\begin{figure}[htb]
\begin{center}
\includegraphics[width=0.45 \textwidth]{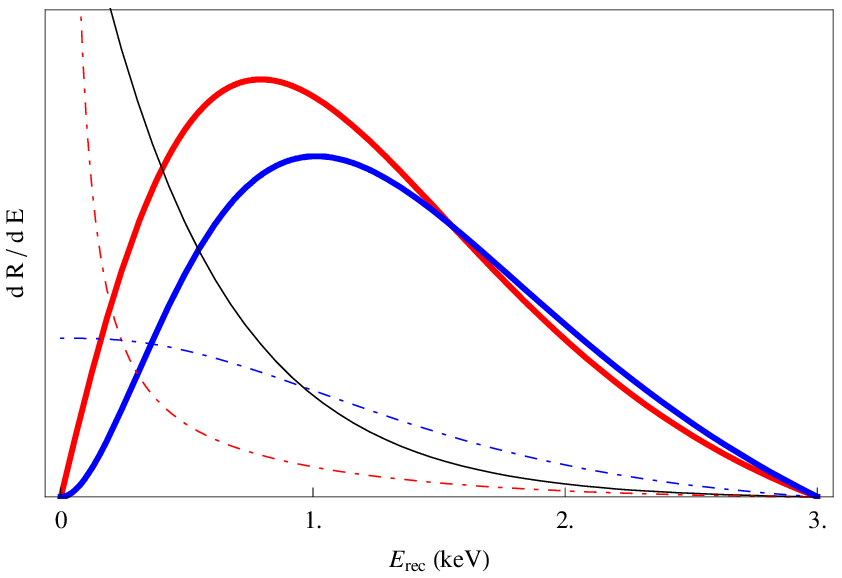}
\includegraphics[width=0.443 \textwidth]{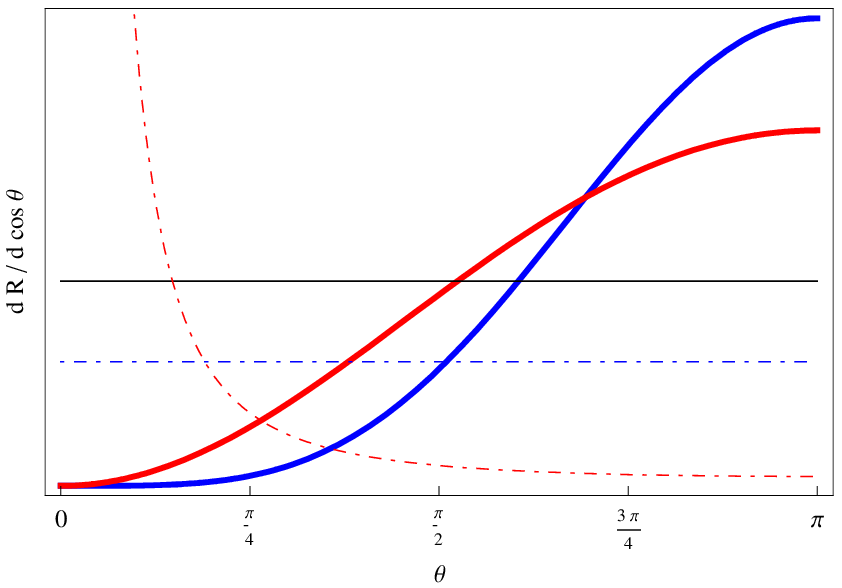}
\caption{These two graphs display the spectral and angular
distribution of SI (blue solid line) and SD (red dashed line)
differential cross sections. The dot-dashed blue (SI) and red (SD)
lines represent the special cases when $m_{\gamma^\prime}=0$,
while the black thin solid line stands for the conventional SI (SD)
interactions. We have chosen dark matter mass to be 5 GeV, $c_2=0$
and used an arbitrary scale in making the above plots.} \label{f2}
\end{center}
\end{figure}

From Eqs.~(\ref{SI0}) and (\ref{SD0}), one can see that the
spectral distribution of the cross sections are quite different
from the conventional SI and SD interactions, as shown in
Fig.~\ref{f2}. We also plot the SI and SD differential cross
sections as a function of the scattering angle $\theta$ in the CM
frame. This is a distinct feature of the new type of interactions
which could be tested in low threshold direction sensitive DM
detectors~\cite{Sciolla:2008vp}.

Before closing this section, we comment that our formulae for the
cross sections can be generalized to the case of a dark matter carrying
magnetic dipole moment that couples directly to the normal photon in the visible sector
as well~\cite{Masso:2009mu}  by simply replacing
$\varepsilon_{\gamma'}^2/m_{\gamma'}^4$ with $1/|{\bf q}|^4$.
In addition, we emphasize that the dimension six operators in
Eq.~(\ref{Lprime}) should also be taken into account for completeness.

\section{Direct detection in an asymmetric DM model }\label{sec3}
We start this section with a few more details about the asymmetric
DM model proposed in~\cite{An:2009vq} in addition to those
outlined in the introduction. The particle masses and symmetry
breaking in the two sectors are generated via the usual Higgs
mechanism. We introduce two Higgs doublets $H^{(\prime)}_{u,d}$ in
both sectors obeying $Z_2$ symmetries so that the up-type fermions
only couple to $H_u$ or $H_u'$; whereas the down-type fermions to
$H_d$ or $H_d'$. This avoids the tree level flavor changing
neutral currents. We add soft mirror symmetry breaking terms,
which may arise from a mirror symmetric model at high scale via
spontaneous symmetry breaking~\cite{bere1}. They allow us to have
symmetry breaking patterns in the two sectors different while the
interactions and associated coupling constants remain symmetric.
This way, one can get the mirror Higgs doublet vacuum expectation
values (VEVs) to be larger than those of the SM Higgs. We can also
break mirror electric charge while keeping the familiar
$U(1)_{em}$ unbroken. In order to implement the inverse seesaw mechanism to
give light neutrino masses, we add two $Y=2$ triplet Higgs fields
to both sectors~\cite{An:2009vq} which acquire different vacuum
expectation values. The asymmetric symmetry breaking pattern has
several consequences:

\begin{itemize}

\item The the dark sector particles are heavier than the SM
particles. Taking the ratio $\tan\beta^\prime\equiv
v^\prime_{u}/v^\prime_d>\tan\beta\equiv v_u/v_d$ and proper
parameters value,\footnote[2]{In particular, we take
$m_u=2.5\,$MeV, $m_d=5\,$MeV, $m_s=98\,$MeV,
$\Lambda_{QCD}=200\,$MeV and $v_{\rm
wk}=\sqrt{v_u^2+v_d^2}=246\,$GeV. We also fix $\tan\beta=50$ in
the visible sector, which means $y_t\simeq y_b$. If the lightest
mirror baryon mass is chosen to be 5\,GeV, different values of
$\tan\beta'$ determines $\Lambda_{QCD}'$, $v_{\rm wk}'$ and the
identity of DM, see Sec.~\ref{darkneutron}} one can have the
mirror neutron as the lightest mirror baryon with mass $\sim 5$
GeV, which then becomes the dark matter candidate.

\item The mirror sector $U(1)_{em}'$ breaking gives the mirror
photon a mass and the kinetic mixing $(\varepsilon_\gamma/2)
F^{\mu\nu}F_{\mu\nu}'$ between the two $U(1)$'s allows the massive
mirror photon to decay into the familiar electron-positron pair.
The lifetime of the mirror photon is
$\tau_{\gamma'}\approx\left({50\,\rm{MeV}}/{m_{\gamma'}}\right)\left
({7\times 10^{-11}}/{\varepsilon_\gamma}\right)^2 \rm{sec} \ . $
For $m_{\gamma'}=50$ MeV, $\varepsilon_\gamma > 7 \times 10^{-11}$
is needed to avoid the constraints from BBN. QED precision
measurements provide constraints on the coupling
$\varepsilon_\gamma$. The most stringent constraint comes from the
measurement of the muon magnetic moment, which gives an upper
bound $\varepsilon_\gamma^2< 2\times 10^{-5} (m_{\gamma'}/100\,
\mbox{MeV})^2$~\cite{Pospelov:2008zw}.

\item The kinetic mixing between the familiar photon with the mirror
photons allows dark matter to directly scatter against nuclei
making direct detection of asymmetric mirror dark matter possible.

\end{itemize}

\subsection{Mirror Neutron as DM}\label{darkneutron}

As discussed in~\cite{An:2009vq}, the mirror neutron can be the lightest
 mirror baryon state and hence qualified to be the dark matter
candidate, provided $\tan\beta'$ lies in the window
$100<\tan\beta'<233$. Here we will choose $\tan\beta'=150$ as an
example and the mirror neutron mass to be $m_{n'}=5\,$GeV.
Correspondingly the next-to lightest mirror baryon, the mirror
proton has a mass $m_{p'}=5.7\,$GeV and mirror $\Delta$-baryon
$m_{\Delta'^-}=5.8\,$GeV, and $\Lambda'_{QCD}=1.1\,$GeV, $v'_{\rm
wk}=210\,$TeV.

At low energies, the mirror neutron dark matter interacts with the
nucleons in the target via the kinetic mixing between the photon
and the mirror photon (of the broken $U(1)'_{e.m.}$) characterized
by the parameter $\varepsilon_\gamma$ . The interaction takes the
general form as we show in Sec.~\ref{sec21}.

The mirror neutron is composed of three mirror quarks with masses
higher than the intrinsic scale of mirror strong interaction.
 To calculate the direct detection rate,
 one must determine the Wilson coefficients
 $c_1$ and $c_2$ in Eq.~(\ref{Lprime}). They are related to the electromagnetic
 form factors of the mirror neutron
\begin{eqnarray}
\bar \chi(p') \left[ F_1(q^2) \gamma^\mu + F_2(q^2) \frac{1}{2 m_\chi} i \sigma^{\mu\nu} q_\nu \right]
 \chi(p) A'_\mu(q) \ ,
\end{eqnarray}
where the mirror electric charge is $F_1(0)=0$ and
\begin{eqnarray}
F_1'(0) = \frac{c_2}{2m_\chi^2}, \ \ \ F_2(0) = 2 c_1 \ .
\end{eqnarray}
Therefore the Wilson coefficients $c_1$ and $c_2$ are related to
the physical quantities of the magnetic dipole moment
and the generalized ``charge radius" as defined in~\cite{Isgur:1998er}
\begin{eqnarray}
\mu_\chi = \frac{c_1 e}{m_\chi}, \ \ \ r^2_{E\chi} = \frac{3(c_1+c_2)}{m_\chi^2}\,.
\end{eqnarray}

Since the mirror nucleon is a composite particle,
we estimate its anomalous magnetic dipole moment by using the naive quark model,
\begin{eqnarray}
\mu_\chi \simeq - \frac{1}{3} \frac{Q_{u'}e}{2 m_{u'}} + \frac{4}{3} \frac{Q_{d'} e}{2 m_{d'}} \ ,
\end{eqnarray}
and thereby fix $c_1$. However, due to the non-perturbative nature, the coefficient
$c_2$ is not easily determined in the same picture.
Therefore, in the following numerical discussions, we will take $c_2$ to be the same
order as $c_1$ as a free parameter.
In this case, we can rewrite Eq.~(\ref{SI0}) for SI interaction in terms of $\mu_\chi$
\beq
\frac{d \sigma_{\rm SI}}{dE_r} =
\varepsilon_\gamma^2
\frac{Z^2 e^2 m_A \mu^2 v^2\mu_\chi^2}{2\pi m_{\gamma'}^4}
\left[\left( 1+ \frac{c_2}{c_1} \right)^2\frac{\mu^2}{m_\chi^2} (1-\cos\theta)^2 + \sin^2 \theta\right] F^2(|{\bf q}|)\,.
\eeq

\begin{figure}[htb]
\begin{center}
\includegraphics[width=0.45 \textwidth]{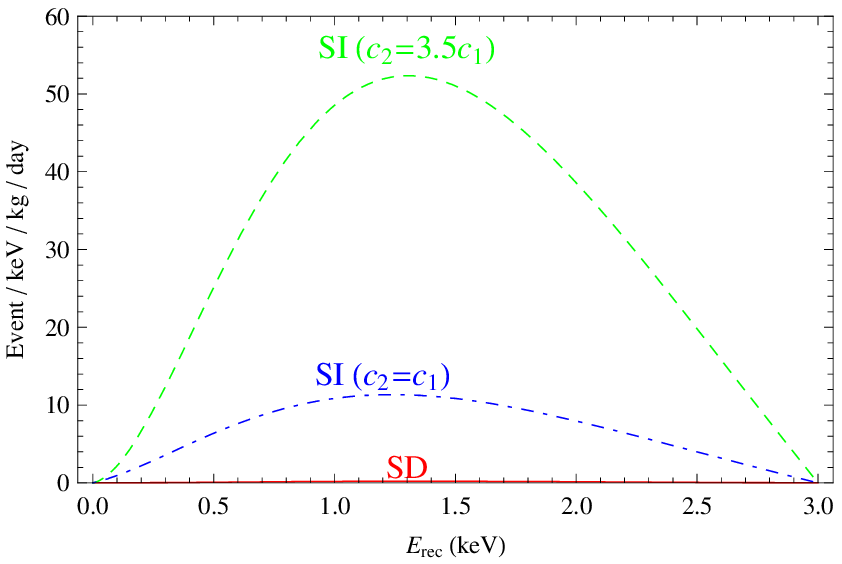}
\includegraphics[width=0.45 \textwidth]{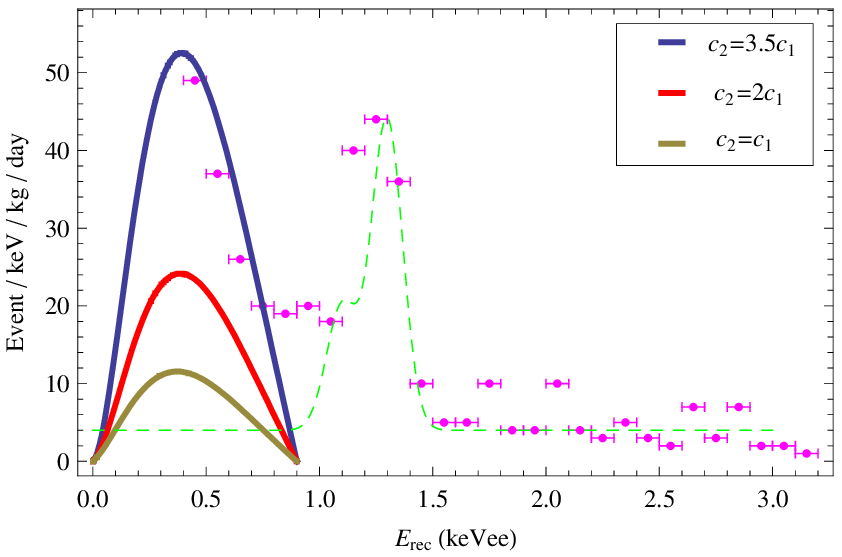}
\caption{The event rate spectral for $^{73}Ge$ target detector
with 5 GeV mirror neutron as the DM. In the left graph, we show the SI and SD contributions separately.
In the right panel, we show a fit for the CoGeNT data with different $c_2$ values. The dashed line
represents the Gaussian peaks from the L-shell background component.}
\label{Cogentfig}
\end{center}
\end{figure}

\medskip
\noindent{\bf CoGeNT}\\
The CoGeNT experiment observed possible dark matter event counts
 between ionization energy $0.4-3.2\,$keVee~\cite{Aalseth:2010vx}.
 While the excesses around $1-1.5\,$keVee are attributed to a
 background component describing the L-shell energy levels associated with
 electron capture in $^{68}$Ge and $^{65}$Zn,
 the first few bins below 1\,keVee can be interpreted to arise out of dark matter scattering~\cite{Chang:2010yk}.
To fit the data, we take $m_{\gamma'}=10\,$MeV and $\varepsilon_\gamma^2=2\times10^{-7}$.
With a light dark matter mass of 5\,GeV, the nuclear form factors are very close to 1~\cite{Servant:2002hb}
and the upper bound on the recoil energy is only a few keV.
The SI and SD contributions to the detection rates are displayed in the left graph
in Fig.~\ref{Cogentfig} for different values of $c_2$, the SI cross section is the dominant contribution.
We also show the total rate spectral with different values of $c_2$ and a quenching factor $Q=0.3$.
As explained before, at very low energy, the event rate tends to vanish instead
 of increasing exponentially, due to the ${\bf q}$-dependent interactions.
 We find that when taking $c_2\approx 3.5c_1$ one can fit the experimental data well.

\begin{figure}[htb]
\begin{center}
\includegraphics[width=0.45 \textwidth]{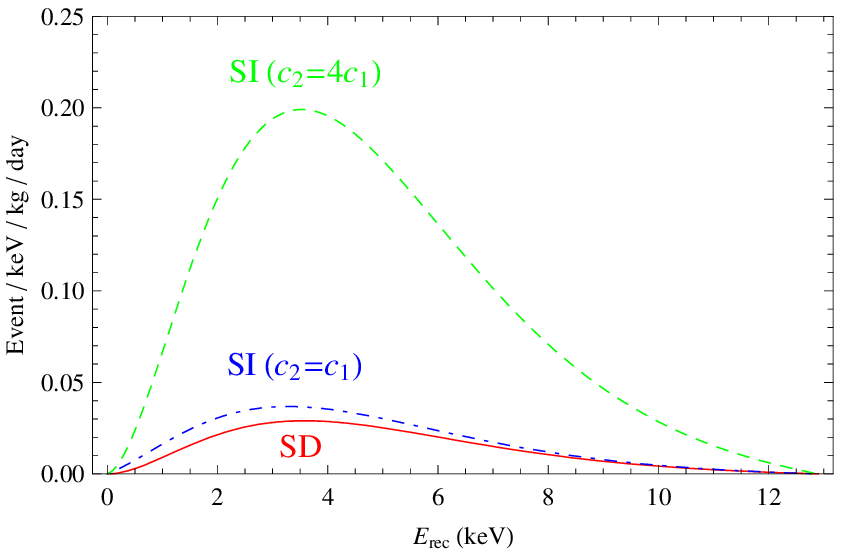}
\includegraphics[width=0.45 \textwidth]{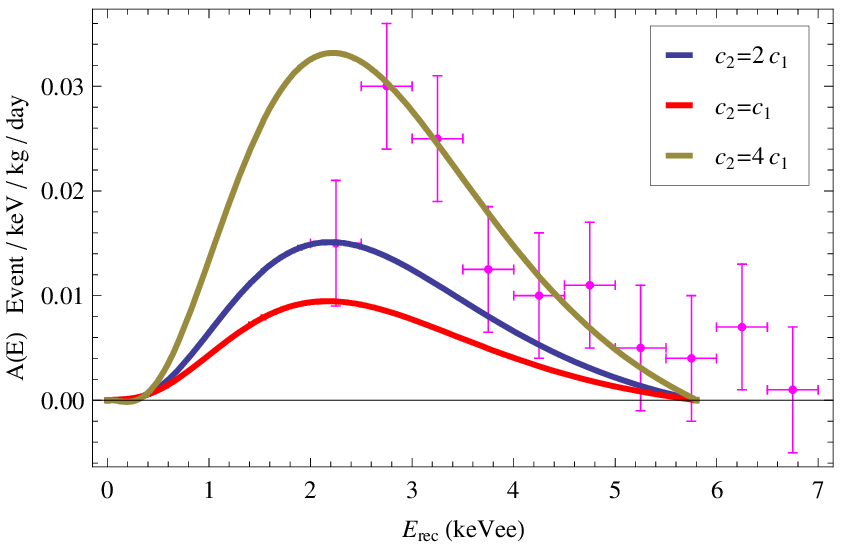}
\caption{The event rate spectral for DAMA and a fit for the annual modulation amplitude $A(E)$.}
\label{DAMAfig}
\end{center}
\end{figure}

\medskip
\noindent{\bf DAMA}\\
The DAMA collaboration has reported an annual modulation signal in the
scintillation light~\cite{Bernabei:2003za,Bernabei:2008yi}.
The scattering of the light dark matter with the sodium nucleus
yields $1-10\,$keVee ionization energy.
Due to the relative small $Z=11$ of the sodium, we find the the SD cross section
is numerically comparable to the SI counterpart, if $c_2\simeq c_1$. This is because although the SI
 contribution is enhanced by a coherent factor $Z^2$, the SD amplitude merits
  a nonrelativistic factor $1/(m_\chi m_p)$ rather than $1/(2m_\chi^2)$ thus gaining
  an extra relative factor of more than $\sim 10^2$ in the cross section.
The total detection rate is not very sensitive to the precise value of $c_2$,
as long as $c_1$ and $c_2$ are of the same order.
To study the annual modulation observed by DAMA, we take the parameter values as
$m_{\gamma'}=10\,$MeV and $\varepsilon_\gamma^2=0.5\times10^{-7}$ and choose
a quenching factor $Q=0.45$. The predicted annual modulation amplitude $A(E)$
is shown in Fig.~\ref{DAMAfig}.

Clearly, with our choice of parameters, simultaneous fits to both
DAMA and CoGENT appears somewhat difficult. Note however that due
to quantum corrections, the primordial mirror lepton asymmetry
could easily be different from the lepton asymmetry in the visible
sector. This could easily allow a dark matter mass higher than 5
GeV, which will improve the simultaneous fits. We feel that at
this stage, it is premature to get into such detailed
phenomenological study.

Second point, we wish to make concerns the recent paper by the
Xenon100 collaboration\cite{x100}, which seems to rule out the
light dark matter region favored by CoGENT and DAMA. There appears
to be some controversy\cite{contro} regarding the results and it
is prudent to wait till situation clears.

\subsection{Mirror Proton or $\Delta'$-Baryon as DM}\label{sec_charged}

For small $\tan\beta'<100$, the mirror proton is lighter than the
mirror neutron and will therefore be the dark matter.  On the
other hand, for very large $\tan\beta'>233$, the mirror
$\Delta'^-$-baryon can be the dark matter candidate. In the quark
model picture, the mirror neutron $n'$ is composed of $(u'd'd')$,
while the mirror $\Delta'^-$ is composed of $(d'd'd')$. In the
large $\tan\beta$ regime, the $m_{d'}<m_{u'}$, so one might
naively expect that $m_{\Delta'^-}<m_{n'}$. However, in the QCD
where the current quark masses are negligible, there is a mass
splitting between the neutron and $\Delta$-baryons, which is about
300 MeV. This mass difference is understood to be due to the
hyperfine interaction among the constituent quarks, which is
proportional to the baryonic wave function at the origin,
\begin{eqnarray}
m_{\Delta}-m_{n} \sim \frac{|\psi_B(0)|^2}{m_q^2} \approx \frac{\Lambda_{QCD}^3}{m_q^2} \ ,
\end{eqnarray}
where $m_q$ is the constituent quark mass $\sim300\,$MeV. For the mirror sector QCD, we
can estimate by using a similar expression with the intrinsic scale and quark masses scaled,
\begin{eqnarray}
\frac{m_{\Delta'^-}-m_{n'}}{m_{\Delta}-m_{n}} \approx \left( \frac{\Lambda_{QCD}'}{\Lambda_{QCD}}\right)^3
\left( \frac{m_q^2}{m_{u'} m_{d'}} \right) \ ,
\end{eqnarray}
where in contrast $m_{u'}$ and $m_{d'}$ are mirror quark current
masses, since they are heavy in the mirror sector.  The hyperfine
interaction tends to compensate the mass difference due to
$m_{u'}>m_{d'}$. Taking $\tan\beta'=300$ and
$m_{\Delta'^-}=5\,$GeV as the DM, we get $m_{p'}=9.6\,$GeV,
$m_{n'}=6.2\,$GeV, $\Lambda_{QCD}'=1.2\,$GeV and $v_{\rm
wk}'=365\,$TeV.

The important point as far as direct detection is concerned is
that both the mirror proton $p'$ and the $\Delta'^{-}$-baryon are
charged under mirror electromagnetism and the interaction with
detector nuclei is not suppressed by $\sim{\bf q}^2/m_N^2$. The
Lagrangian for the interaction of dark matter with the mirror
photon is given by
\begin{eqnarray}
\mathcal{L}' = c_0 e \bar \chi \gamma^\mu \chi A_\mu' \ ,
\end{eqnarray}
where $c_0=\pm1$ represents the mirror electric charge of dark matter.

This gives the conventional SI cross section with \beq \sigma_{\rm
SI}^0 = \varepsilon_\gamma^2 \frac{Z^2 c_0^2 e^4 \mu^2}{\pi
m_{\gamma'}^4} \,. \eeq Taking a model-allowed value
$\varepsilon_\gamma = 6\times  10^{-8} (m_{\gamma^\prime} /50{\rm
MeV})^2$, one can obtain the cross section per nucleon
$\sigma_{\chi N} \simeq \sigma^0_{\rm SI}  \mu_{\chi N}^2
/(\mu_{\chi A}^2 A^2) \simeq 7\times 10^{-41} {\rm cm^2}$, which
is required to account for the events observed by the CoGeNT
collaboration.

\section{Summary}\label{sec4}
To summarize, we have presented a general operator analysis of an
asymmetric dark matter interacting with nucleons via a mirror photon
and applied it to an asymmetric mirror dark matter model suggested
by four of us in a previous paper. We note that when the dark
matter is neutral under dark electromagnetic forces (zero mirror
electric charge), {\it e.g.}, mirror neutron, it interacts with
nucleons via the mirror magnetic dipole moment and electric charge
radius. In this case, there is an energy dependence in the direct
detection cross section
 as well as an angular dependence different from the usual massive symmetric WIMP case
  ({\em e.g.}, SUSY case). As the sensitivities of dark matter searches improve, one can use
   these results to pinpoint the detailed nature of dark matter interaction with matter.

\section*{Acknowledgments}
We acknowledge Bobby Acharya and Xiangdong Ji for useful discussions.
This work was partially supported by the U. S. Department of Energy via grant DE-FG02-
93ER-40762. The works of RNM is supported by the NSF grant PHY-0652363.

\end{document}